\documentclass[aps,prl,twocolumn,superscriptaddress,showpacs,floatfix]{revtex4-1}

\usepackage{bm}
\usepackage{amsmath}
\usepackage{amsfonts}
\usepackage{amssymb}
\usepackage{longtable}
\usepackage{epsf}
\usepackage[dvips]{epsfig,color}
\usepackage{version}
\usepackage{graphicx}
\usepackage{psfrag}

\newcommand{\be}{\begin{equation}}
\newcommand{\ee}{\end{equation}}

\begin{document}
\title{Experimental evidence of enhanced transport in supernematics}

\author{Wycliffe K. Kipnusu}
\author{Philipp Zeigermann}
\author{Ciprian Iacob}
\author{Friedrich Kremer}
\affiliation{Institute of Experimental Physics I, University of Leipzig, 04103
Leipzig, Germany}
\author{Martin Schoen}
\affiliation{Stranski-Laboratorium f\"ur Physikalische und Theoretische Chemie,
Technische Universit\"at Berlin, Stra{\ss}e des 17. Juni 135, 10623 Berlin, Germany}
\affiliation{Department of Chemical and Biomolecular Engineering, Engineering
  Building I, Box 7905, North Carolina State University, 911 Partners Way,
  Raleigh, NC 27695, USA}
\author{Marco G. Mazza}
\affiliation{Max Planck Institute for Dynamics and Self-Organization, G\"{o}ttingen, Germany}
\author{Rustem Valiullin}
\affiliation{Institute of Experimental Physics I, University of Leipzig, 04103
Leipzig, Germany}

\date{\today}
\begin{abstract}
  Deviations of molecular shapes from spherical symmetry may give rise to a variety of novel phenomena, including their dynamic behavior. It has recently been predicted [Mazza \textit{et al}. Phys. Rev. Lett. \textbf{105}, 227802 (2010)] that liquid crystals in the reentrant nematic phases may show unexpectedly high rates of translational displacements. Using broadband dielectric spectroscopy and a single-component thermotropic liquid crystal we explore molecular dynamics in the vicinity of the reentrant nematic transition and find the formation of a maximum in the mobilities upon changing temperature. The occurrence of the high mobility states are found to be history-dependent due to the formation of metastable mesophases on heating. The experimental results are further supported by computer simulation. Our results contribute to the clarification of an ongoing controversy on the dynamics of reentrant nematics.

\end{abstract}
\maketitle

The diffusive transport of molecules plays a key role in various phenomena including cellular metabolism,
crystal growth, catalysis \cite{Heitjans2005}.  These processes typically occur at (quasi)equilibrium conditions, hence their control is hardly achievable without severely affecting  the external conditions. Finding a means by which the diffusion processes can be favorably altered is therefore a challenging fundamental problem with direct implications for many potential applications. One of the approaches for transport control is based on the exploitation of coupling between the phase state and dynamics. In the same way, mesostructure-dynamics correlations in materials with asymmetric interactions leading to structural transitions can be exploited to control transport  without affecting the phase state. Such an asymmetry can be provided by the molecular shapes \cite{Han2006, Fakhri2010, Frise2010, Turiv2013, Chakrabarty2013}. The most prominent and thoroughly addressed examples are liquid crystals (LC) in which an ample structural behavior is translated into respective rich transport properties.

In some cases, the structure-dynamics correlations may lead to very striking and counterintuitive phenomena. For example, Goto \emph{et al}.~showed increase of charge mobility in the nematic phase of LCs \cite{Goto2007, Goto2010}. Recently, other computer simulation studies revealed that non-polar LCs, exhibiting a reentrant nematic (RN) phase, can show enhanced rates of translational mobility \cite{Mazza2010,Mazza2011,Mazza-rev2011,Stieger2012}. This finding was attributed to the lowering of the rotational configurational entropy upon forming a denser RN phase. It was further anticipated that, under this condition, the LC molecules may more easily move laterally in a collective way in a form of string-chains.

Experimental verification of this phenomenon has remained a challenging task. The study by Dvinskikh and Fur\'{o} \cite{Dvinskikh2012} is to the best of our knowledge the only one in which this problem was directly addressed. They measured the rates of translational mobility along the nematic axis in a LC mixture showing reentrant behavior using the pulsed field gradient (PFG) technique of NMR, but found no indications of an enhanced dynamics in the RN-phase. Notably, they used a mixture of 6OCB and 8OCB LCs, which, at certain mixing fractions, exhibits the RN phase. It may therefore be possible that the occurrence of a high mobile state in this work was prohibited by the peculiar structure of the mesogens of two types. Another reason could be strong dipolar interactions of the mesogens studied. In fact, many contradictory results are found about the dynamics of reentrant nematics (see \cite{Mazza-rev2011} for a review). In order to avoid the aforementioned problems, in the present work we have used a single-component nematogen 4-cyanophenyl-3'-methyl-4(4'-n-dodecylbenzoyloxy)benzoate (12 CPMBB). This LC is known to exhibit a sequence of isotropic (I) - nematic (N) - smectic A (smA) - reentrant nematic (RN) phases upon
temperature variation with the transition temperatures $T_{IN}=420.9$ K,  $T_{NA}=411.6$ K, and $T_{ARN}=332.9$ K, respectively \cite{Ratna1981}.

To probe the LC dynamics, we used broadband dielectric spectroscopy (BDS). The measurements were performed using a Novocontrol high-resolution alpha analyzer ($10^{-1}$-$10^7$ Hz) and Hewlett Packard (HP) 4291A impedance analyzer ($10^7$-$10^9$ Hz) under pure nitrogen atmosphere in broad frequency ($10^{-1}$-$10^9$ Hz) and temperature ($260$-$430$ K) ranges. The LC was prepared in a form of thin film with a thickness of about 100 $\mu$m. The thus prepared film was confined between two platinum electrodes. It has been already proven that, by probing fluctuations in the orientations of dipolar molecules, this technique is capable of delivering the rates of molecular translational propagations \cite{Iacob2008, Sangoro2008, Sangoro2011}. In this way, the diffusivities of different molecular species were probed in a broad range from $10^{-10}$ to $10^{-20}$ m$^{2}$/s and were shown to be in a good agreement with the results obtained using other experimental techniques, in particular by PFG NMR.

\begin{figure}
\includegraphics[scale=0.34]{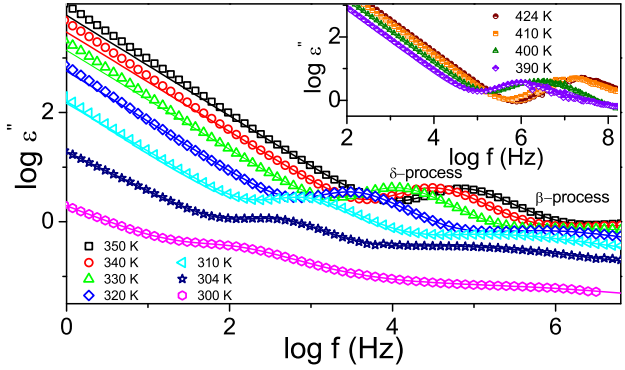}
\caption{Dielectric loss versus frequency for the selected temperatures. Lines represent fits using Eq.~\ref{HN} and the conductivity contribution. The inset includes data obtained at high frequencies up to 10$^{8}$ Hz. The error bars are smaller than the symbols. \label{epsilon_relax}}
\end{figure}

Figure~\ref{epsilon_relax} shows the primarily measured dielectric loss spectra of the LC for a few selected temperatures. The frequency dependence of the complex permittivity was fitted by the empirical Havriliak-Negami function \cite{Kremer2003}
\be
\varepsilon^*_{HN}(\omega)=\varepsilon_\infty+\frac{\Delta\varepsilon}{\left[1+(i\omega\tau_{HN})^\alpha\right]^\gamma}+\frac{\sigma_0}{i\omega\varepsilon_0}
\label{HN}
\ee
where $\Delta\varepsilon$ is the dielectric relaxation strength, $\tau_{HN}$ is the characteristic time constant, which is related to relaxation time at maximum loss, $\sigma_0$ is the dc-conductivity, $\varepsilon_0$ is the permittivity of free space, and $0< \gamma, \gamma\alpha \leq 1$ represent the broadening of the loss peaks. Two relaxation processes (Fig.~\ref{epsilon_relax}) contribute to the dielectric loss. The $\delta$-process is assigned to the hindered rotational (libration) motion around the short axis of the molecule, i.e. associated with the fluctuations of the dipolar moment-component perpendicular to the molecular axis. Conversely, the component parallel to this axis gives rise to the $\beta$-process, which corresponds to librational fluctuations around the long molecular axis.

\begin{figure}
\begin{center}
\includegraphics[scale=0.32]{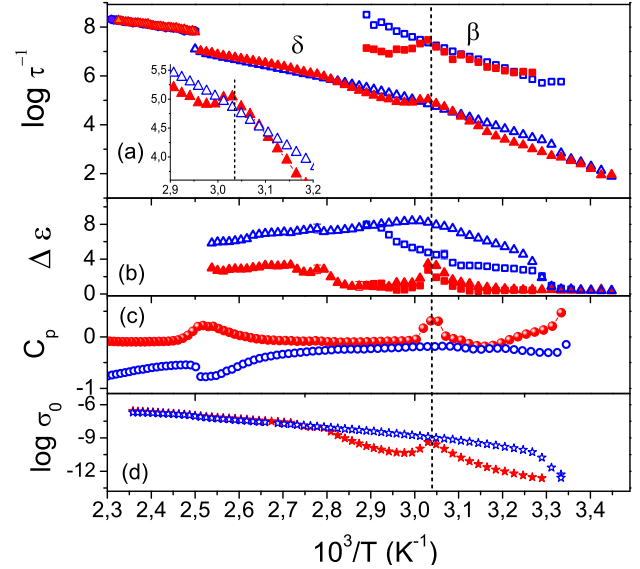}
\end{center}
\caption{(a) Dielectric relaxation rates ($\delta$-process (up triangles) and $\beta$-process (squares)) in $s^{-1}$, (b) dielectric relaxation strengths, (c) specific heat capacity in $Jg^{-1}K^{-1}$, and (d) dc-conductivity in $Scm^{-1}$ versus inverse temperature. The inset in (a) shows zoomed in part of $\delta$-process around $T=329$ K, which is shown by vertical lines. Filled (red) and open (blue) symbols denote heating and cooling runs, respectively.  \label{rates_relax}}
\end{figure}

The activation plot of the relaxation rates $\tau^{-1}$, obtained from the fits for the $\beta-$ and $\delta$-processes, are shown in Fig.~\ref{rates_relax}a. Only a single relaxation process was observed in the isotropic phase, which splits into the $\delta-$ and $\beta-$processes for the ordered phases of the LC. After the characteristic jump at the isotropic-nematic transition, a smooth decrease of the relaxation rates is found upon further cooling, except a small downhill step at $\sim 302$ K which may be associated with the crystallization transition. Importantly, the results were proven to be invariant upon changing the cooling rate. These findings correlate with the differential scanning calorimetry (DSC) data of Fig.~\ref{rates_relax}c. On cooling, characteristic peaks are observed at the same temperatures. As the most important feature, however, the behavior changes dramatically on heating. The DSC shows a typical peak in the heat capacity at the RN phase transition temperature $T_{ARN}=329$ K, which is close to $T_{ARN}$ reported earlier \cite{Ratna1981}. Most remarkably, an enhancement of the relaxation rates for both $\beta$- and $\delta$-processes are noted to occur around $T_{ARN}$.

\begin{figure}
\includegraphics[scale=0.34]{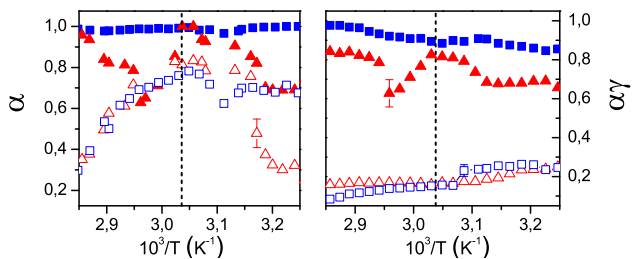}
\caption{The Havriliak-Negami distribution shape parameters $\alpha$ and $\alpha\gamma$ vs inverse temperature for the $\delta$- (filled symbols) and $\beta$-process (open symbols) on cooling (squares) and heating (triangles). The vertical line show $T=329$ K. \label{AG}}
\end{figure}

The specific changes of the relaxation rates observed on cooling and heating (Fig.~\ref{rates_relax}) are accompanied by a peculiar behavior of the relaxation time distributions (RTD). For the Havriliak-Negami model, RTD may have an asymmetric shape, which is described by two parameters \cite{Sinha1998}. Thus, $\alpha$ quantifies broadening of the low-frequency branch of RTD, while the product $\alpha\gamma$ is responsible for the high-frequency part. These two quantities are shown in Fig.~\ref{AG}. Most spectacular behavior is observed for the $\delta$-process. On cooling, $\alpha$ and $\alpha\gamma$ are both close to 1.0 revealing almost Debye-like relaxation, i.e. very narrow RTD. Both low- and high-frequency parts of RTD broaden dramatically with the onset of crystallization around $300$ K. Upon heating, RTD remains notably broader than on cooling and reveals some patterns, which are absent on cooling. Most notably, both $\alpha$ and $\alpha\gamma$ show peaks around $T_{ARN}$ with the magnitudes nearly identical to those obtained on cooling. Because the $\delta$-process corresponds to fluctuations about the molecular long-axis, the peaks for the $\delta$-process at $T_{ARN}$ can be associated with the decrease of orientational fluctuations observed in MD simulations and proposed as the driving mechanism for enhanced transport in the RN phase \cite{Mazza2010}.

The hysteresis observed in experiments is often associated with the occurrence of metastable states. In order to examine their occurrence, we have performed an experiment in which the sample was first cooled down to only $T=320$ K (above the onset of crystallization) and then the relaxation rates were measured during heating. In this case, no maxima in the relaxation rates were observed (not shown). This renders the crystallization process crucial for the development of the hysteresis, revealing the emergence of metastabilities during heating from the crystalline state. We associate them with the formation of mesoscalic domains with different order axes during melting. The thus created domains relax to a common orientation only at sufficiently high temperatures.

This scenario is supported by the temperature dependencies of the dielectric strength $\Delta \epsilon$ shown in Fig.~\ref{rates_relax}b. The data on $\Delta \epsilon$, which is most sensitive to the structural changes occurring in the samples \cite{Kremer2003, Stocchero2004}, reveal two important observations. First of all, the maxima in $\Delta \epsilon$ close to $T_{ARN}$ indicate the occurrence of structural reorganizations both on cooling and heating. Secondly, upon heating from 280 K, i.e. from the crystalline state, the peak in $\Delta \epsilon$ is found to be very sharp, but with a lower amplitude than on cooling. Noting that $\Delta \epsilon$ is proportional to $\mu_{x}^{2}$, where $\mu_{x}$ is the component of the molecular polarisability along the electric field $\overrightarrow{E}$, this fact may support the hypothesis of the mesoscopic disorder present on heating. This line of reasoning implies that, on the cooling branch, the LC molecules are preferentially orientated with their long axes parallel to $\overrightarrow{E}$ (the latter is perpendicular to the LC film surface in our setup). On heating, $\Delta \epsilon$ results as an average over isotropic orientation of local order axes,  yielding, thus, substantially lower $\Delta \epsilon$. Since the dielectric relaxation and $\Delta \epsilon$ are coupled by the Kramers-Kronig relations \cite{Kremer2003}, the formation of the peaks in the relaxation rates and $\Delta \epsilon$ is attributed to the same physical origin.

\begin{figure}
\includegraphics[width=0.45\textwidth]{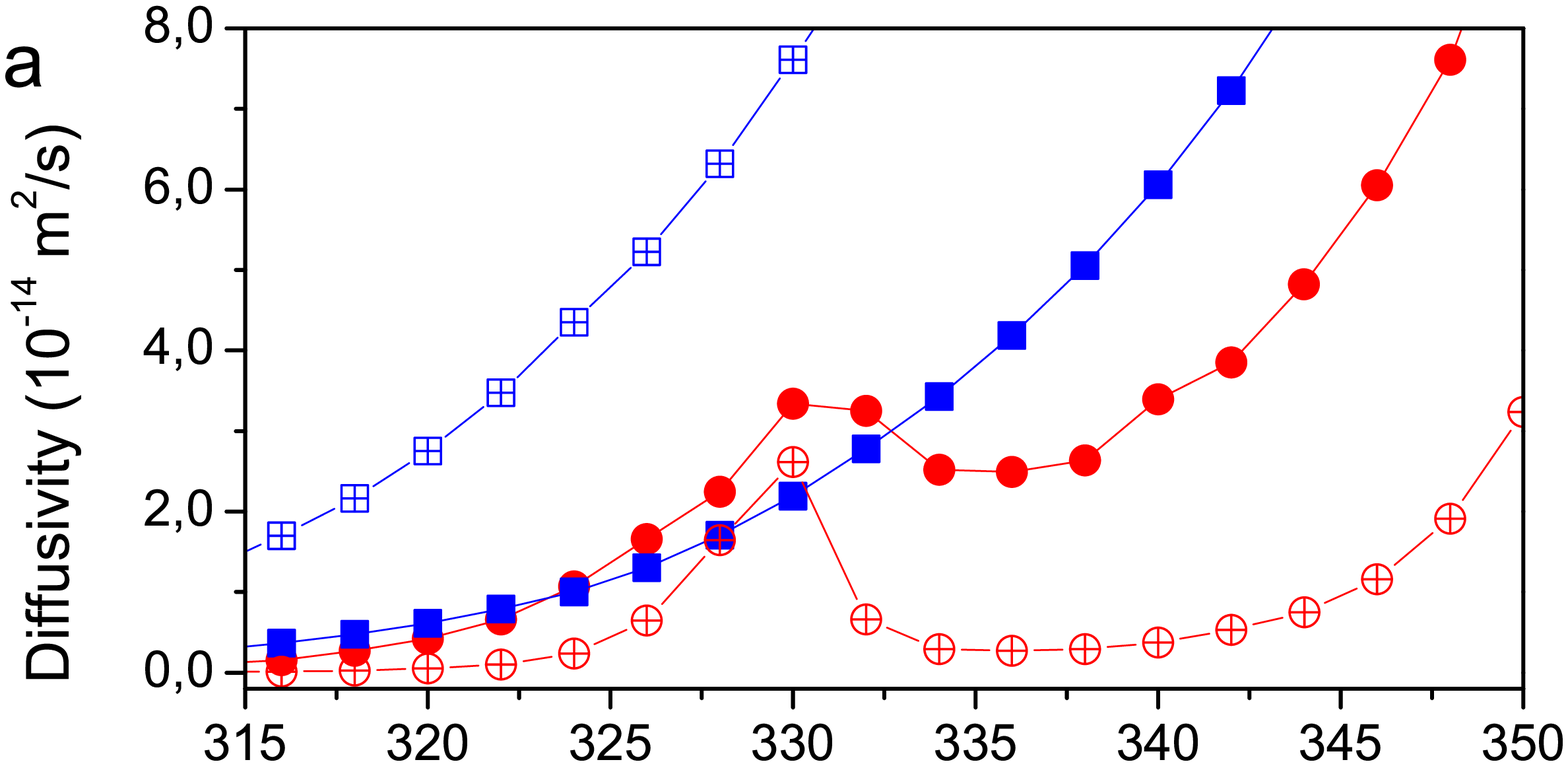}
\includegraphics[width=0.45\textwidth]{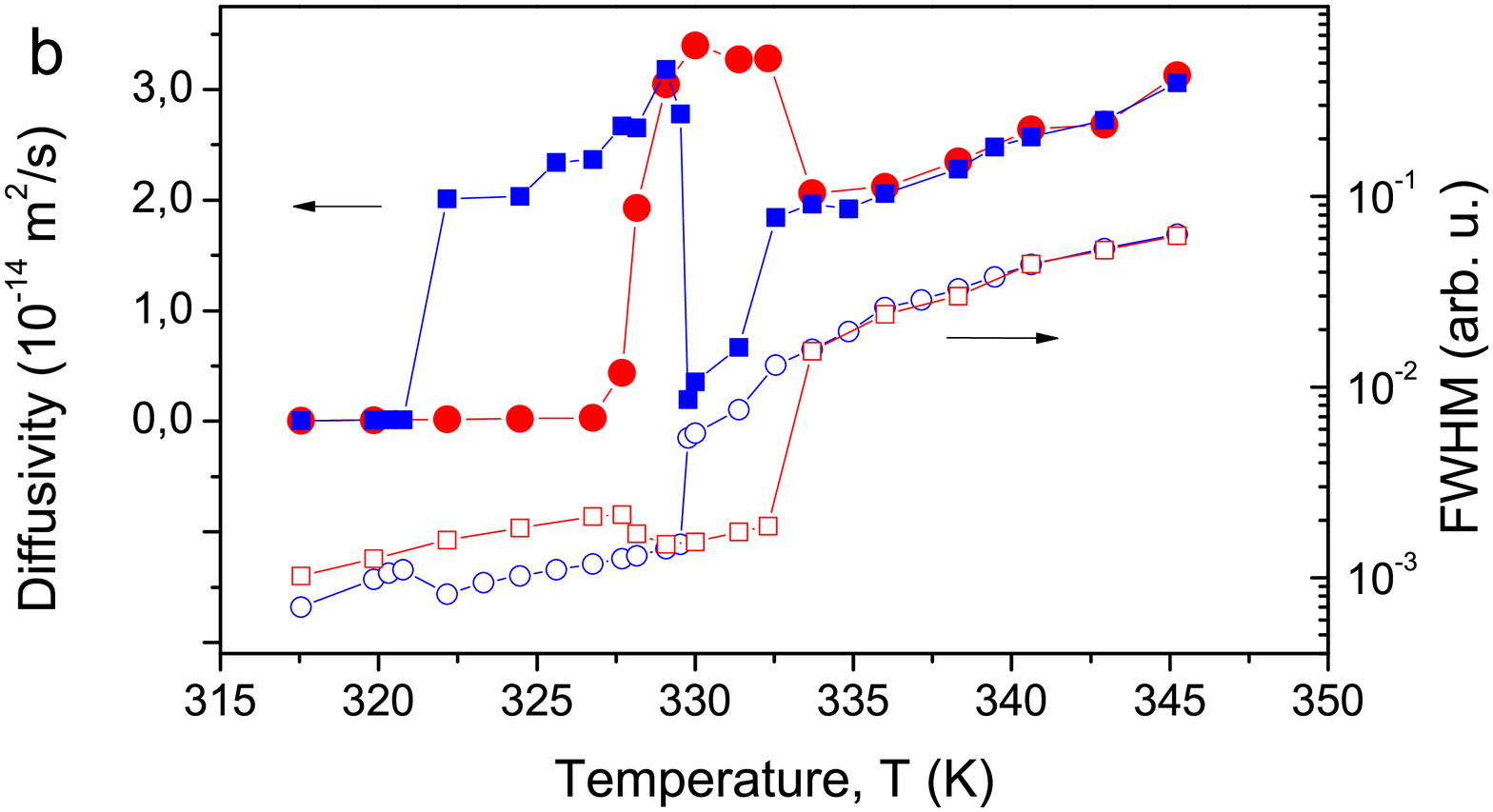}
\caption{Temperature dependence of the diffusivity (a) experimentally obtained from the dielectric relaxation (filled symbols) and conductivity (crossed symbols) and (b) calculated from computer simulations \cite{note1}. The open symbols in (b) show FWHM of the orientational distribution function. The rectangles and circles refer to the cooling and heating branches, respectively.\label{diff}}
\end{figure}

The relaxation data reveal the enhanced rates of the local order fluctuations in the RN-phase attained on heating. In what follows, we are going to correlate them with the rates of molecular translations. Quite generally, $\tau$ can be related to the viscoelastic and dynamic properties of a material as
\begin{equation}
    \tau= \left[ q^{2} \left( \frac{K}{\eta} + D \right ) \right ]^{-1},
    \label{tauq}
\end{equation}
where $q$ is the wave-vector, $K$ is the elastic constant, $\eta$ is the viscosity, and $D$ is the diffusivity. The second term on the right-hand side of Eq.~\ref{tauq} takes into account the orientation fluctuations caused by molecular translations \cite{Kimmich1997}. Under the condition of $K/\eta < D$, $\tau$ is determined by the orientational fluctuations due to diffusion. In this respect, only the $\delta$-process can plausibly be associated with the displacement of the center of mass of the LC molecules, while this is unlikely for the $\beta$-process. Therefore, we may relate the $\delta$-relaxation rates with the diffusivity via the Einstein-Smoluchowski relation
$D=\lambda^{2} / 2 \tau$, where $\lambda$ is the mean jump length. To assess $\lambda$, we independently measured $D$ and $\tau$ in the isotropic phase of LC using PFG NMR and DS, respectively \cite{Sangoro2011}. Thereafter, the thus obtained $\lambda$ was used to obtain the diffusivities in the ordered LC phases using the DS data for the relaxation rates. The resulting $D$ are shown in Fig.~\ref{diff}a for temperatures around $T_{ARN}$.

To provide an independent support for the conjecture that the enhanced relaxation rates result from the enhanced molecular mobilities, we additionally measured the dc-conductivities $\sigma_{0}$ (see Fig.~\ref{rates_relax}d). Notably, charge transfer in the LC studied is largely determined by the mass transfer. By using the Nernst-Einstein equation, the diffusivities can be estimated from $\sigma_{0}$. The respective data are also shown in Fig.~\ref{diff}a. For the transformation, the charge density was estimated using the independently measured diffusivity (using PFG NMR) and electric conductivity in the isotropic LC phase. Fig.~\ref{diff}a reveals that the estimates of the diffusivities obtained from the dielectric relaxation and conductivity measurements show an enhancement of the diffusivity on heating around $T_{ARN}$.

All the experimental results obtained unveil the following picture. Upon cooling from the isotropic state, the LC molecules assume a common axis below $T_{IN}$. With further decrease of the temperature, the RN-phase is not formed due to metastability. Upon heating from the crystalline state, a large number of mesoscopic domains with differing local order axes emerge. At $T_{ARN}$, the occurrence of the RN-phase is manifested. The phase is characterized by a narrow distribution of the relaxation rates, which presumably results from a low orientational entropy. The latter is a prerequisite for the diffusivity enhancement according to the model suggested in Ref.~\cite{Mazza2010}.

To validate this scenario, we have also performed molecular dynamics (MD) simulation of an apolar LC using the Gay-Berne-Kihara (GBK) model \cite{martinez04} with the same model parameters as in Ref.~\cite{Mazza2010}.  We simulated a system of $N=4000$ molecules in the isothermal-isobaric ensemble  confined in the $z$-direction by walls at fixed distance $s_{\mathrm{z}}=20\sigma$, where $\sigma$ is the LC diameter, to facilitate the formation of ordered phases at lower pressures; periodic boundary conditions are used in the $x$- and $y$-directions. To accommodate for the LC ordered phase, we allowed independent changes of the simulation box in the $x$- and $y$-directions. Calculation of physical quantities is then performed in the isothermal-isochoric ensemble to have a minimal perturbation of the dynamics. Molecular interactions are cut off beyond $d_{\mathrm{c}}=3$.  Similar to the experiments, we consider a cooling and a heating sequence of simulations, where the molecular configuration equilibrated at one $T$ is used as the initial configuration to be equilibrated at the next $T$ in the sequence.

We calculate the molecular mean square displacement (MSD)
\begin{equation}\label{eq:msd}
\left\langle
\big[\Delta r\left(t\right)\big]^2
\right\rangle
\equiv\frac{1}{N}\left\langle
\sum\limits_{i=1}^{N}\left[\bm{r}_i\left(t_0+t\right)-\bm{r}_i\left(t_0\right)
\right]^2\right\rangle
\end{equation}
where $\langle\cdot\rangle$ means ensemble average and average over initial times $t_0$. From the long time behavior of the MSD we extract the value of diffusivity $D$ according to Einstein's relation
\begin{equation}
D=\lim\limits_{t\to\infty}
\frac{1}{6t}
\left\langle
\big[\Delta r\left(t\right)\big]^2
\right\rangle
\end{equation}
In addition, we calculate the orientational distribution function $f(\cos(\theta))$, where $\theta$ is the angle between the molecular orientations and the global director.

Figure~\ref{diff}b shows the $T$-dependence of $D$ for an isobaric cooling and heating. Starting from $T=345$~K in the cooling sequence, $D$ has an initial steady decrease from the isotropic phase into the nematic phase. The abrupt dip is due to the smectic phase, which has a particularly small diffusivity.  At $T\approx 330$~K there is a sudden increase of $D$ as the system enters the RN phase. $D$ decreases smoothly again until the transition to a solid crystal, occurring at $T=320$~K. In the heating sequence the system is in a crystalline solid phase at $T\approx 317$~K. As $T$ is raised there is a prominent jump in $D$ at $T$ slightly above $327$~K as the system enters the RN phase. Importantly, the jump in the diffusivity is nicely correlated with the respective decrease of the orientational distribution width. We note the marked shift to higher $T$ than in the cooling sequence, which is a clear indication of hysteresis. At $T\approx 337$~K the fluid enters the nematic phase (the smectic phase is prevented due to metastability), displaying a jump to lower values of $D$. From this point on, $D$ increases monotonically and reach the same values as in the cooling sequence. Finally, the comparison of Fig.~\ref{diff}a and \ref{diff}b show qualitative agreement between theory and experiment.

In conclusion, by means of dielectric spectroscopy and electric conductivity measurements we have studied the molecular transport in a single-component liquid crystal exhibiting the reentrancy phenomenon, i.e. the formation of the RN phase. We show the enhancement of molecular mobilities in this phase prepared by heating from the crystalline state in accord with the predictions made earlier using molecular dynamics simulations \cite{Mazza2010}. The occurrence of the high mobility states in the RN phase is found to be accompanied by the narrowing of the distribution of the dielectric relaxation rates, presumably revealing lowering of the orientational entropy, which is observed in the MD simulations. The results presented here point to a direct correlation between orientational entropy and diffusion for single-component RNs. The fundamental connection between entropy and diffusion was proposed by Dzugutov for simple liquids almost $20$ years \cite{Dzugutov1996}. Now we provide a similar relationship for liquid crystals.

The work is supported by the German Science Foundation (DFG) and the Max Planck Society. M.S. is grateful for financial support from the International Graduate Research Training Group 1524, R.V. and F.K. from the Research Group FOR 877. We gratefully acknowledge Michaela Dzionara for providing us with the LC sample.



\begin{thebibliography}{10}
\bibitem{note1}%
  \BibitemOpen
  \bibinfo {note} {To rescale the temperature in reduced units, $T^*$ (used in
  MD simulations), to SI units, $T$, we apply a linear scaling,
  $T=330+4.61(T^*-4.7)$, obtained by matching the abscissae of the minimum and
  maximum of diffusivity in the experimental heating curve to the corresponding
  points in the MD heating curve. To rescale the diffusivity in reduced units,
  $D^*$, to that in SI units, $D$, we use a logarithmic rescaling,
  $\ln{D}-\ln{D_1}=\frac{\ln{D_2}-\ln{D_1}}{\ln{D_2^*}-\ln{D_1^*}} ( \ln{D^*} -
  \ln{D_1^* })$ such that $\ln{D}=1.24\ln{D^*}+2.17$. The latter choice assumes
  that the dynamics is exponentially slower at low temperatures.}%
  \bibAnnoteFile{Stop}{note1}%
\end{thebibliography}

%

\end{document}